\documentstyle[aps,preprint]{revtex}

\begin{document}

\title{Maximum Mass-Radius Ratio for Compact General Relativistic Objects
in Schwarzschild- de Sitter Geometry}
\author{M. K. Mak\footnote{E-mail:mkmak@vtc.edu.hk} and Peter N. Dobson, Jr.\footnote{E-mail:aadobson@ust.hk}}
\address{Department of Physics, The Hong Kong University of Science and Technology,\\
Clear Water Bay, Hong Kong, P. R. China.}
\author{T. Harko\footnote{E-mail:tcharko@hkusua.hku.hk}}
\address{Department of Physics, The University of Hong Kong,
Pokfulam Road, Hong Kong, P. R. China.}

\maketitle

\begin{abstract}
Upper limits for the mass-radius ratio are derived for
arbitrary general relativistic matter distributions in the presence
of a cosmological constant. General restrictions for the red shift and total energy
(including the gravitational contribution) for compact objects
in the Schwarzschild-de Sitter geometry are also
obtained in terms of the cosmological constant and of the mean
density of the star.
\end{abstract}


Pacs Numbers: 04.20.-q, 97.60.-s

\narrowtext

\section{Introduction}

The possibility that the cosmological constant be nonzero and dominates the
energy density of the Universe today is one of the most intriguing problems
of the contemporary physics. Data recently collected by two survey teams
(the Supernova Cosmology Project \lbrack 1\rbrack and the High-$z$ Supernova Search Team \lbrack 2\rbrack)
and analyzed in the framework of
homogeneous FLRW cosmological models, have yielded, as a primary result, a
strictly positive cosmological constant of order unity.

Within the classical GR, the existence of a cosmological constant is
equivalent to the postulate that the total energy momentum tensor of the
Universe $T_{ik}^{\left( U\right) }$ possesses an additional piece $%
T_{ik}^{(V)}$, besides that of its matter content $T_{ik}^{\left( m\right) }$%
, of the form $T_{ik}^{(V)}=\Lambda g_{ik}$, where generally the
cosmological constant $\Lambda $ is a scalar function of space and time.
Such a form of the additional piece has previously been obtained in certain
field - theoretical models and is interpreted as a vacuum contribution to
the energy momentum tensor \lbrack 3-5\rbrack, $\Lambda g_{ik}=\left\langle
T_{ik}^{(V)}\right\rangle =\frac{8\pi G}{c^{4}}\left\langle \rho
_{V}\right\rangle g_{ik}$ , where $\rho _{V}$ is the energy of the vacuum.
The vacuum value of $T_{ik}$ thus appears in the form of a cosmological
constant in the gravitational field equations (for a review of the cosmological constant problem see \lbrack 6\rbrack).

The cosmological constant can also be interpreted as a parameter measuring
the intrinsic temperature of the empty space-time, or in a sense, of the
geometry itself \lbrack 7\rbrack. 

At interplanetary distances, the effect of the cosmological constant could
be imperceptible. However, Cardona and Tejeiro \lbrack 8\rbrack\ have shown
that a bound of this constant can be obtained using the values observed from
the Mercury's perihelion shift in a Schwarzschild - de Sitter space-time.
The presence of a cosmological constant implies that, for the first time
after inflation, in the present epoch its role in the dynamics of the
Universe becomes dominant. On the other hand there is the possibility that
scalar fields presents in the early Universe could condense to form the so
called boson stars \lbrack 9-10\rbrack. There are also
suggestions that the dark matter could be made up of bosonic particles. This
bosonic matter would condense through some sort of Jeans instability to form
compact gravitating objects. A boson star can have a mass comparable to that
of a neutron star \lbrack 11\rbrack. The simplest kind of boson star is
made up of a self - interacting complex scalar field $\Phi $ describing a
state of zero temperature \lbrack 12-13\rbrack. The
self-consistent coupling of the scalar field to its own gravitational field
is via the Lagrangian $L=\frac{1}{2k}\sqrt{-g}R+\frac{1}{2}\sqrt{-g}\left[
g^{ik}\Phi _{,i}^{\ast }\Phi _{,k}-V\left( \left| \Phi \right| ^{2}\right)
\right] $ where $V\left( \left| \Phi \right| ^{2}\right) $ is the
self-interaction potential usually taken in the form $V\left( \left| \Phi
\right| ^{2}\right) =\frac{1}{2}m^{2}\left| \Phi \right| ^{2}+\frac{1}{4}%
\lambda \left| \Phi \right| ^{4}$. $m$ is the mass of the scalar field
particle (the boson) and $\lambda $ is the self-interaction parameter. For
the bosonic field the stationarity ansatz is assumed, $\Phi (t,r)=\phi
(r)e^{-i\omega t}$. If we suppose that in the star's interior regions and
for some field configurations the scalar field is constant then in the
gravitational field equations the scalar field self-interaction potential will play the role of a
cosmological constant, which could also describe a mixture of ordinary matter
and bosonic particles.

By using the static spherically symmetric gravitational field
equations Buchdahl \lbrack 14\rbrack\ has obtained an absolute constraint of
the maximally allowable mass $M$- radius $R$ ratio for isotropic fluid
spheres of the form $\frac{2M}{R}<\frac{8}{9}$ (we use
natural units $c=G=1$). It is the purpose of the present Letter to
investigate the maximum allowable mass -radius ratio in the case of compact
general relativistic objects in the presence of a cosmological constant and
to study the possible effects of the cosmological constant upon the
red-shift and total energy of general relativistic compact objects.

\section{Maximum mass-radius ratio for compact objects in Schwarzschild- de Sitter geometry}

For a static general relativistic spherically symmetric matter configuration
with interior line element given by $ds^{2}=e^{\nu }dt^{2}-e^{\lambda
}dr^{2}-r^{2}\left( d\theta ^{2}+\sin ^{2}\theta d\varphi ^{2}\right) $ the
components of the energy-momentum tensor are $T_{0}^{0}=\rho $, $%
T_{1}^{1}=T_{2}^{2}=T_{3}^{3}=-p$, where $\rho $ is the energy density and $%
p $ the thermodynamic pressure.

In the presence of a cosmological constant the properties of a compact object 
 can be described completely by the gravitational structure
equations, which are given by: 
\begin{equation}\label{1}
\frac{dm}{dr}=4\pi \rho r^{2},
\end{equation}
\begin{equation}\label{2}
\frac{dp}{dr}=-\frac{\left( \rho +p\right) \left[ m+4\pi \left( p-\frac{%
2\Lambda }{3}\right) r^{3}\right] }{r^{2}\left( 1-\frac{2m}{r}-\frac{8\pi }{3%
}\Lambda r^{2}\right) },  
\end{equation}
\begin{equation}\label{3}
\frac{d\nu }{dr}=\frac{2\left[ m+4\pi \left( p-\frac{2\Lambda }{3}\right)
r^{3}\right] }{r^{2}\left( 1-\frac{2m}{r}-\frac{8\pi }{3}\Lambda
r^{2}\right) },  
\end{equation}
where $m(r)$ is the mass inside radius $r$ .

Eqs. (\ref{1})-(\ref{3}) must be considered together with an equation of state
of the dense matter, $p=p(\rho )$ and with the boundary conditions $%
p(R)=0 $, $p(0)=p_{c}$ and $\rho (0)=\rho _{c}$, where $\rho _{c}$ and $p_{c}
$ are the central density and pressure, respectively.

With the use of Eqs. (\ref{1})-(\ref{3}) it is easy to show that the function $\zeta =e^{%
\frac{\nu }{2}}>0,\forall r\in \lbrack 0,R\rbrack $ obeys the equation 
\begin{equation}\label{4}
\sqrt{1-\frac{2m(r)}{r}-\frac{8\pi }{3}\Lambda r^{2}}\frac{1}{r}\frac{d}{dr}%
\left[ \sqrt{1-\frac{2m(r)}{r}-\frac{8\pi }{3}\Lambda r^{2}}\frac{1}{r}\frac{%
d\zeta }{dr}\right] =\frac{\zeta }{r}\frac{d}{dr}\frac{m(r)}{r^{3}}.
\end{equation}

Since the density $\rho $ does not increase with increasing $r$, the mean
density of the matter $<\rho >=\frac{3m(r)}{4\pi r^{3}}$ inside radius $r$
does not increase either. Therefore we assume that inside a compact general
relativistic object the condition $\frac{d}{dr}\frac{m(r)}{r^{3}}<0$ holds
independently of the equation of state of dense matter.

Introducing a new independent variable $\xi =\int_{0}^{r}r^{\prime }\left( 1-%
\frac{2m(r^{\prime })}{r^{\prime }}-\frac{8\pi }{3}\Lambda r^{\prime
2}\right) ^{-\frac{1}{2}}dr^{\prime }$ \lbrack 13\rbrack, from Eq.(\ref{4}) we
obtain the basic result that all stellar type general relativistic matter
distributions with negative density gradient obey the condition 
\begin{equation}\label{5}
\frac{d^{2}e^{\frac{\nu \left( \xi \right) }{2}}}{d\xi ^{2}}<0,\forall r\in
\left[ 0,R\right].
\end{equation}

Using the mean value theorem we conclude $\frac{de^{\frac{\nu \left( \xi
\right) }{2}}}{d\xi }\leq \frac{e^{\frac{\nu \left( \xi \right) }{2}}-e^{%
\frac{\nu \left( 0\right) }{2}}}{\xi }$, or, taking into account that $e^{%
\frac{\nu \left( 0\right) }{2}}>0$ we find $\frac{de^{\frac{\nu \left( \xi
\right) }{2}}}{d\xi }\leq \frac{e^{\frac{\nu \left( \xi \right) }{2}}}{\xi }$%
. In the initial variables we have 
\begin{equation}\label{6}
\frac{m\left( r\right) +4\pi \left( p-\frac{2\Lambda }{3}\right) r^{3}}{r^{3}%
\sqrt{1-\frac{2m}{r}-\frac{8\pi }{3}\Lambda r^{2}}}\leq \left[
\int_{0}^{r}r^{\prime }\left( 1-\frac{2m(r^{\prime })}{r^{\prime }}-\frac{%
8\pi }{3}\Lambda r^{\prime 2}\right) ^{-\frac{1}{2}}dr^{\prime }\right]
^{-1}.  
\end{equation}

Since for stable stellar type compact objects $\frac{m}{r^{3}}$ does not
increase outwards $\frac{m(r^{\prime })}{r^{\prime }}\geq \frac{m(r)}{r}%
\left( \frac{r^{\prime }}{r}\right) ^{2}$, $\forall r^{\prime }\leq r$. \ We
denote $\alpha \left( r\right) =1+\frac{4\pi }{3}\Lambda \frac{r^{3}}{m(r)}$
. Moreover, we assume that in the presence of a cosmological constant the
condition $\frac{\alpha \left( r^{\prime }\right) m\left( r^{\prime }\right) 
}{r^{\prime }}\geq \frac{\alpha \left( r\right) m\left( r\right) }{r}\left( 
\frac{r^{\prime }}{r}\right) ^{2}$, or, equivalently, 
\begin{equation}\label{7}
\left( 1+\frac{4\pi }{3}\Lambda \frac{r^{\prime 3}}{m\left( r^{\prime
}\right) }\right) \frac{m\left( r^{\prime }\right) }{r^{\prime }}\geq \left(
1+\frac{4\pi }{3}\Lambda \frac{r^{3}}{m\left( r\right) }\right) \frac{%
m\left( r\right) }{r}\left( \frac{r^{\prime }}{r}\right) ^{2},  
\end{equation}
holds inside the compact object. In fact Eq. (\ref{7}) is independent  
of the cosmological constant $\Lambda $ and is valid for all decreasing density compact
matter distributions.

Therefore we can evaluate the RHS of Eq. (\ref{6}) as follows: 
\begin{eqnarray}\label{8}
&\int_{0}^{r}\frac{r^{\prime }}{\left( 1-\frac{2m\left( r^{\prime }\right) 
}{r^{\prime }}-\frac{8\pi }{3}\Lambda r^{\prime 2}\right) ^{\frac{1}{2}}}%
dr^{\prime }=\int_{0}^{r}\frac{r^{\prime }}{\left( 1-\frac{2\alpha \left(
r^{\prime }\right) m\left( r^{\prime }\right) }{r^{\prime }}\right) ^{\frac{1%
}{2}}}dr^{\prime }\geq   \nonumber\\
&\int_{0}^{r}\frac{r^{\prime }}{\left[ 1-\frac{2\alpha \left( r\right) m(r)}{r%
}\left( \frac{r^{\prime }}{r}\right) ^{2}\right] ^{\frac{1}{2}}}dr^{\prime }
=\frac{r^{3}}{2\alpha \left( r\right) m(r)}\left[ 1-\left( 1-\frac{2\alpha
\left( r\right) m(r)}{r}\right) ^{\frac{1}{2}}\right] \nonumber\\
\end{eqnarray}

With the use of Eq. (\ref{8}), Eq.(\ref{6}) becomes: 
\begin{equation}\label{9}
\frac{m\left( r\right)+4\pi \left( p-\frac{2\Lambda }{3}\right) r^{3}}{%
\sqrt{1-\frac{2m(r)}{r}-\frac{8\pi }{3}\Lambda r^{2}}}\leq \frac{2m(r)\left(
1+\frac{4\pi }{3}\Lambda \frac{r^{3}}{m(r)}\right) }{1-\sqrt{1-\frac{2m(r)}{r%
}-\frac{8\pi }{3}\Lambda r^{2}}}.
\end{equation}

Eq. (\ref{9}) is valid for all $r$ inside the star. It does not depend on
the sign of $\Lambda $.

Consider first the case $\Lambda =0$. By evaluating (\ref{9}) for $r=R$ 
we obtain $\frac{1}{\sqrt{1-\frac{2M}{R}}}\leq 2\left[ 1-\left( 1-\frac{2M}{R%
}\right) ^{\frac{1}{2}}\right] ^{-1}$, leading to the well-known result $%
\frac{2M}{R}\leq \frac{8}{9}$ \lbrack 13\rbrack .

For $\Lambda \neq 0$, Eq. (\ref{9}) leads to the following upper limit for
mass-radius ratio of compact objects 
\begin{equation}\label{10}
\frac{2M}{R}\leq \left( 1-\frac{8\pi }{3}\Lambda R^{2}\right) \left[ 1-\frac{%
1}{9}\frac{\left( 1-\frac{2\Lambda }{\bar{\rho}}\right) ^{2}}{1-\frac{8\pi 
}{3}\Lambda R^{2}}\right],  
\end{equation}
where $\bar{\rho}=\frac{3M}{4\pi R^{3}}$ is the mean density of the star.

In order to find a general restriction for $\bar{\rho}$ we shall consider
the behavior of the Ricci invariant $r_{2}=R_{ijkl}R^{ijkl}$. If the static line
element is regular, satisfying the conditions $e^{\nu (0)}=const.\neq 0$ and 
$e^{\lambda (0)}=1$, then the Ricci invariants are also non-singular
functions throughout the star. In particular for a regular space-time the
invariants are non-vanishing at the origin $r=0$. For the invariant $r_{2}$
we find 
\begin{equation}\label{11}
r_{2}=\left( 8\pi \rho +8\pi p-\frac{4m}{r^{3}}\right) ^{2}+2\left( 8\pi
p-8\pi \Lambda +\frac{2m}{r^{3}}\right) ^{2}+2\left( 8\pi \rho +8\pi \Lambda
-\frac{2m}{r^{3}}\right) ^{2}+4\left( \frac{2m}{r^{3}}\right) ^{2}.
\end{equation}

For a monotonically decreasing and regular pressure and density functions, the
function $r_{2}$ is also regular and monotonically decreasing throughout the
star. Therefore it satisfies the condition $r_{2}(R)<r_{2}(0)$. By assuming that the
surface density is vanishing, $\rho (R)=\rho _{surface}=0$, we obtain 
the following general constraint upon the mean density of the star: 
\begin{equation}\label{12}
\left( \frac{2M}{R^{3}}\right) ^{2}+2\left( \frac{M}{R^{3}}-4\pi \Lambda
\right) ^{2}<8\pi ^{2}\left[ \left( p_{c}+\frac{\rho _{c}}{3}\right)
^{2}+2\left( p_{c}+\frac{\rho _{c}}{3}-\Lambda \right) ^{2}+2\left( \frac{%
2\rho _{c}}{3}+\Lambda \right) ^{2}+\frac{4}{9}\rho _{c}^{2}\right].
\end{equation}

\section{Discussions and Final Remarks}

The existence of a limiting value of the mass-radius ratio leads to upper
bounds for other physical quantities of observational interest. One of these quantities is the surface red shift $z$%
, defined in the Schwarzschild-de Sitter geometry according to $z=\left( 1-%
\frac{2M}{R}-\frac{8\pi }{3}\Lambda R^{2}\right) ^{-\frac{1}{2}}-1$.

In the absence of a cosmological constant Eq.(\ref{9}) leads to the
well-known constraint $z\leq 2$. For $\Lambda \neq 0$ the surface red shift
must obey the general restriction 
\begin{equation}\label{13}
z\leq \frac{3}{1-\frac{2\Lambda }{\bar{\rho}}}-1.
\end{equation}

As another application of the obtained upper mass -radius ratios we shall
derive an explicit limit for the total energy of the compact general
relativistic star. The total energy (including the gravitational field
contribution) inside an equipotential surface $S$ can be defined to be
\lbrack 14\rbrack\ 
\begin{equation}\label{14}
E=E_{M}+E_{F}=\frac{1}{8\pi }\xi _{s}\int_{S}\left[ K\right] dS,
\end{equation}
where $\xi ^{i}$ is a Killing field of time translation, $\xi _{s}$ its
value at $S$ and $\left[ K\right] $ is the jump across the shell of the
trace of the extrinsic curvature of $S$, considered as embedded in the
2-space $t=constant$. $E_{M}=\int_{S}T_{i}^{k}\xi ^{i}\sqrt{-g}dS_{k}$ and $%
E_{F}$ are the energy of the matter and of the gravitational field,
respectively. This definition is manifestly coordinate invariant. In the
case of a static spherically symmetric matter distribution from Eq. (\ref{14}%
) we obtain the following exact expression: $E=-re^{\frac{\nu -\lambda }{2}}$
\lbrack 14\rbrack . Hence the total energy of a compact general relativistic
object in a Schwarzschild-de Sitter space-time is $E=-R\left( 1-\frac{2M}{R}-%
\frac{8\pi }{3}\Lambda R^{2}\right) $.

For $\Lambda =0$ we find the following upper limit for the total energy of
the star: $E\leq -\frac{R}{9}$. In the presence of a cosmological constant
we have 
\begin{equation}\label{15}
E\leq -\frac{R}{9\left( 1-\frac{2\Lambda }{\bar{\rho}}\right) ^{2}}
\end{equation}

In the present Letter we have considered the mass-radius ratio bound for
compact general relativistic objects in the more general Schwarzschild- de
Sitter geometry. Also in this case it is possible to obtain explicit
inequalities involving $\frac{2M}{R}$ as an explicit function of the
cosmological constant $\Lambda $. The surface red shift and the total energy
(including the gravitational one) are modified due to the presence of the
extra term in the gravitational field equations.

\end{document}